# A Low Overhead Reachability Guaranteed Dynamic Route Discovery Mechanism for Dense MANETs

[1]Sharmila Sankar and [2]Dr. V. Sankaranarayanan

[1]Department of Computer Science, B S A University, Chennai, India
sharmilasankar@yahoo.com
[2]Director (University Project), B S A University, Chennai, India
sankarammu@yahoo.com

## ABSTRACT

*A crucial issue for a mobile ad hoc network is the handling of a large number of nodes. As more nodes join the mobile ad hoc network, contention and congestion are more likely. The on demand routing protocols which broadcasts control packets to discover routes to the destination nodes, generate a high number of broadcast packets in a larger networks causing contention and collision. We propose an efficient route discovery protocol, which reduces the number of broadcast packet, using controlled flooding technique. The simulation results show that the proposed probabilistic flooding decreases the number of control packets floating in the network during route discovery phase, without lowering the success ratio of path discoveries. Furthermore, the proposed method adapts to the normal network conditions. The results show that up to 70% of control packet traffic is saved in route discovery phase when the network is denser.*

## 1. INTRODUCTION

In wireless ad hoc networks, the nodes communicate without the aid of any infrastructure. There are many challenges involved in the design of these networks. One particular challenge is involved with the routing of data packets. Typically, the source and the destination nodes for a particular data packet are not within direct communication range. This leads to a multihop scenario where the packets must be routed and forwarded through the other nodes in the network on the way to the destination. Many routing algorithms, like those found in [1-4], have been proposed for ad hoc networks.

In real networks, nodes may join and leave, some (or all) nodes are highly mobile, and node-to-node channels are subject to strong fading. In such cases, the problem of finding routes between given source and destination nodes can present significant difficulties. In particular, there are situations when nodes have to resort to broadcasting. This causes the effect known as "broadcast storm" in large networks, which has been studied in literature [5-9]. Under certain conditions the route discovery process can consume significant portion of network resources and becomes detrimental to overall network performance and stability. For example, if more route discovery processes are initiated by different sources than can be sustained, then they will likely to fail resulting in more retransmissions. In this scenario, the network can become inundated with route request packets and the overall network throughput can significantly decrease.

## 2. FACTORS AFFECTING ROUTING PERFORMANCE

Various factors like Link capacity, Link and node capability, network density, etc. affects the performance of the network. The main factors addressed in this paper are the following.

## 2.1. Network Scaling

Scalability can be broadly defined as whether the network is able to provide an acceptable level of service even in the presence of large number of nodes in the network. It is one of the most important open issues of ad hoc networks. Firstly, ad hoc networks suffer, by nature, from the scalability problems in capacity. In a non-cooperative network, where Omni-directional antennas are being used, the throughput decreases at a rate N, where N is the number of nodes [10]. That is, in a network with 100 nodes, a single device gets approximately one tenth of the theoretical data rate of the network interface card at the maximum. This problem, however, cannot be solved except by physical layer improvements, such as smart antennas.

Routing protocols also set some limits for the scalability of ad hoc networks. Route acquisition and service locations are examples of task that will require considerable overhead, which will grow rapidly with the network size. Proactive routing is not applicable in a dense and dynamic environment due to huge amount of broadcast message of topology changes. Reactive protocols allow deploying large networks in the expense of increased route acquisition latency. Demands for shorter latencies for route acquisition limit the network size drastically.

## 2.2. Flooding

Broadcast (diffusion of a message from a source node to all nodes in the network) is a common operation in ad-hoc networks, and it is used by several routing protocols. Flooding (also called blind broadcast) is the simplest broadcast protocol: each node rebroadcasts the message once and discards duplicates. AODV, SLS, GSR, DSR and HSLS use flooding with various improvements (usually by changing the TTL value of the broadcast packet to limit propagation in the network). The flooding approach is reliable but has a high overhead for the routing protocol (in term of number of packets and MAC layer access) and the number of collisions dramatically increases in the case of dense networks [18].

# 3. RELATED WORK

One of the earliest broadcast mechanisms is flooding, where every node in the network retransmits a message to its neighbors upon receiving it for the first time. Although flooding is very simple and easy to implement, it can be very costly and may lead to a serious problem, often known as the broadcast storm problem [11] that is characterized by high redundant packet retransmissions, network contention and collision. Ni et. al. [12] have studied the flooding protocol analytically and experimentally. Their obtained results have indicated that rebroadcasts could provide at most 61% additional coverage and only 41% additional coverage on average over that already covered by previous transmissions. Therefore, rebroadcasts are very costly and should be used with caution.

In [11], Williams et al. have classified the broadcasting techniques into the following four categories: simple flooding, probability-based, area-based, and neighbor knowledge scheme. In the flooding scheme, every node retransmits its neighbors as a response to every newly received packet. The probability-based scheme is a simple way of controlling message floods. Each node rebroadcasts with a predefined probability $p$ [5]. Obviously when $p=1$ this scheme resembles simple (blind) flooding. In the area based scheme, a node determines whether to rebroadcast a packet or not by calculating and using its additional coverage area. Of these, of interest in this study is the probabilistic scheme family of variants. In this category of broadcasting techniques, a mobile node rebroadcasts packets according to a certain probability.

Zhang and Dharma [13] have described a dynamic probabilistic scheme. They use a combination of probabilistic and counter-based approaches. The value of a packet counter does not necessarily correspond to the exact number of neighbors from the current host, since some of its neighbors may have suppressed their rebroadcasts according to their local rebroadcast probability. On the other hand, the decision to rebroadcast is made after a random delay, which increases latency.

## 4. PROPOSED ROUTE DISCOVERY ALGORITHM

### 4.1. Neighborhood Vector Construction

Every node sends an HELLO message when it is up in the network to every other single hop neighbor. This enables the receiving node to populate its neighborhood vector. The neighbor nodes in turn respond with the HELLO message to enable the newcomer to populate its neighborhood vector. The neighborhood vector is thus constructed with local broadcast of HELLO messages between set of mobile nodes.

Local broadcast of HELLO message is triggered periodically in order to have knowledge of the current network topology. Upon receiving the HELLO message, the node computes the distance of the node that transmitted the HELLO message using the Received Signal Strength Indicator (RSSI) and updates the neighborhood vector. The distance between the two nodes is computed using equation (1) shown below:

$d_{km} = antilog_{10}[\{L\text{-}32.45\text{-}20\log_{10}(f)\}/20]$ ……. equation(1)

where

L ( Path Loss) = $P_t – P_r$

   $P_t$ – Power of the transmitter

   $P_r$ – Received Signal Strength (as indicated by RSSI)

f (Radio frequency) = 2.4 GHz

The neighbor nodes of a node are recorded in the neighborhood table in decreasing order of their distances. This enables the selection of farthest neighbors for rebroadcasting route request messages during route discovery phase. Eventually the number of hops a route request message travels is reduced thus constructing a shortest route between a pair of nodes.

### 4.2. Route Discovery

When a node wants to communicate with another node in the network a unique communication path is established between the sender and the receiver nodes. The source node scans the neighborhood vector for the destination. If the destination node is identified to be the single hop neighbor of the source, the source nodes starts transmitting data packets. The transmission of data will be uninterrupted until there is no change in the geographical positions of the source and the destination nodes.

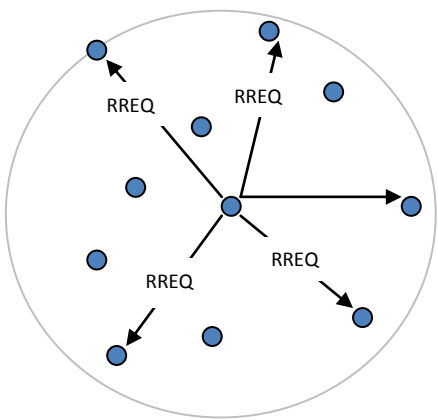

Figure 1. Controlled RREQ broadcast in route discovery

The neighbors in the neighborhood vector are stored in the increasing order of their distances. The source node generates a RREQ packet and forwards it to n/k neighbors (where n is the total number of the neighbors and $K^1$ – reachability parameter - a random number between 3 and 7) from the neighborhood vector targeting the farthest nodes from the source node [14][15](Figure1). The intended neighbors check their neighborhood vectors and locate the destination else the same procedure is repeated till the destination is located. The algorithm (Figure 2) explains the method of finding path from the source to the destination mobile host.

## 4.3. Determination of Number of Rebroadcast

The number of rebroadcasts is determined by the reachability parameter K which ranges between 3 and 7. The number of route requests to be rebroadcasted by each node to determine an optimal path depends on the chosen reachability parameter and the local density of the network. Selecting half of the neighbors from the neighborhood vector in a dense network establishes a shortest path between the source and the destination nodes reducing the control overhead to half from the one that is actually required. The main observations made from the simulations are

- For a less dense network the reachability parameter K should be lesser (i.e) choose at least half of the neighbors from the neighborhood list for rebroadcast.

- For a denser network, the reachability parameter K should be higher for better performance.

This infers that the reachability parameter K is directly proportional to the density D of the network[19]. Due to higher connectivity, which is the inherent characteristics of a dense network, choosing even very few nodes to rebroadcast, discovers a path to the destination, which is closer to optimal path discovered by other broadcast protocols.

*Protocol PathDiscovery()*
*{ while(1)*
  *{ for each chosen neighbor*
      *if destination is in the neighborhood(nbr) vector*
          *{ unicast RREQ to the neighbor;*
            *exit();*
          *}*
      *Choose a value in K such that (3≤ K ≤ 7)*
      *if ((n/k) <= 1)*
          *{ if (n/2 <= 1)*
                *{*
                    *Choose the only neighbor to rebroadcast ;*
                    *Unblock  the neighbor if it was previously blocked by other node in the network*
                *}*
          *}*
      *Choose n/K farthest neighbors from nbr vector;*
      *Block the other neighbors from rebroadcasting;*
      *PathDiscovery();*
  *}*
*}*

Figure 2. Path Discovery Algorithm

---

[1] K divides number of neighbors of a node to choose the candidate for rebroadcasting RREQ. If K is very less many of the neighbors are chosen and only very few are blocked from rebroadcasting. If K is large many neighbors are blocked and only few neighbors rebroadcast.

## 4.4. System Model and Assumptions

We consider a wireless ad hoc network of N nodes with same computation and transmission capabilities, communicating through bidirectional links between each other. In addition we assume a CSMA/CA – MAC layer protocol that provides handshake sequence for control and data transmissions. The information unit for the protocol is the message. It can include data packet as well as control packets.

## 4.5. Simulation Parameters

We carried out the simulation in the customized event driven simulator, OMNET++[16], which is an object modular network test-bed in C++. The mobility scenarios are obtained through mobility framework which is a part of OMNET++ distribution. The scenario generator produces the different mobility patterns such as Random Walk, Random Direction, Random Waypoint entity mobility models. The mobility model chosen for our simulation was Restricted Random Walk. The proposed method was implemented with various densities like 50, 75, 100 and 125 nodes in a terrain region of 350sqm. We compare the number of RREQ rebroadcast in AODV routing protocol with the probabilistic routing model. The MAC layer protocol IEEE 802.11 is used in simulation with the data rate 11Mbps. The data traffic source is set to be a Constant Bit Rate (CBR) source. The network contains one source and one destination, each message packet size of 512 bytes is defined. The Table I provide all the simulation parameter values.

Table1. Simulation Setup Parameters

| Map Size | 350m * 350m |
|---|---|
| Channel Bandwidth | 11 Mbps |
| Channel Delay | 10μsec |
| Simulation Time | 900s |
| Number of Hosts | 50,75,100,125 |
| Channel Gain | 0 |
| Mobility Model | Random Way Point |
| Message Packet Size | 512bytes |

## 4.6 Routing Metrics

Performance of the various protocols depends upon the routing metrics [17]. The main routing metric on the basis of which the performance of the proposed protocol can be determined are:

Saved Rebroadcast (SRB): SRB is defined by (r-t)/r, where r is the number of nodes that received the broadcast message and t is the number of nodes that actually transmitted the message. This evaluates the efficiency and scalability of the routing protocol.

Route Request Success Rate: The percentage of successfully established routes among all route requests.

Route Request Delay: It is defined as the period between the moment when a route request is sent and when a route reply is received.

## 5. RESULTS AND DISCUSSIONS

The simulation studies that were carried out are aimed at evaluating and comparing the performance achieved by Probabilistic Routing Protocol (PRP) and by the other routing protocol under analysis (AODV), during the Route Discovery process. The objective is to show that even after choosing only few neighbor nodes to rebroadcast our protocol performs very well in Dense MANETs.

## 5.1 Simulation Model

Three performance metrics are of interest: (1) Route Request Success Rate, (2) Saved Rebroadcast, and (3) Route Request Delay. In each simulation, we compare the performance of our PRP protocol with AODV protocol, which is well-known and commonly used ad hoc on demand routing protocol.

We vary two system settings: (1) Number of nodes in the network and (2) number of traffic flows in the network to investigate the impacts on the three performance metrics mentioned above.

## 5.2 Determination of Reachability Parameter

The reachability parameter K is an efficiency parameter to achieve the reachability of the broadcast. A very small value in k allows a larger rebroadcast to be flooded in the network and a very large value may suppress many nodes from rebroadcasting, which in turn affects the route discovery. So an extensive study through simulation had been performed to determine a valid threshold for K.

Figure 3 represents the performance of Probabilistic routing method discussed [Section 4.3]. Here the parameter K is a constant and the size of the network is varied between 50 and 125. If the reachability parameter is chosen to be 1 then the protocol behaves as that of AODV. In AODV the number of rebroadcast saved is only 10 to 15% with single source – destination pair. The parameter K used is very useful for partial broadcast. It diffuses the information to a part of the nodes independent of the density.

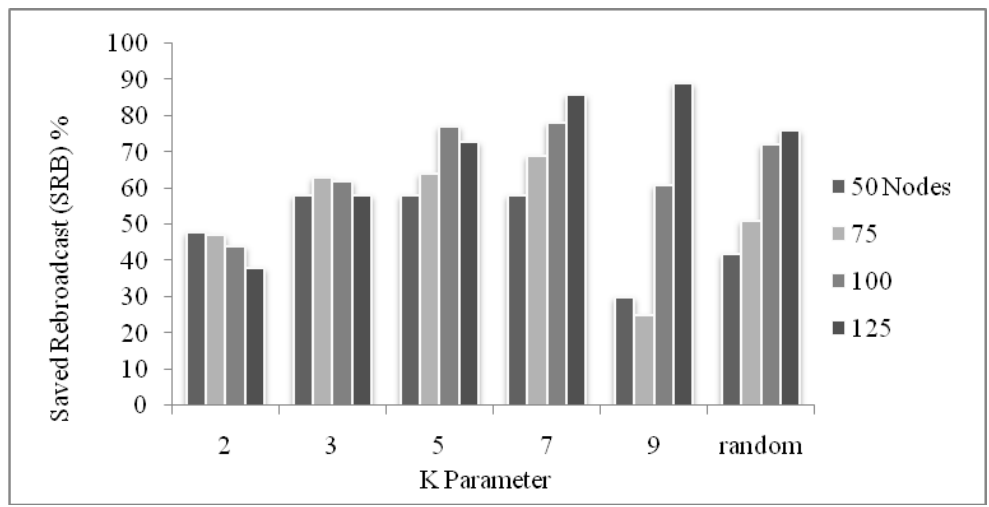

K- ReachabilityParameter
SRB - 2, 3, 5, 7, 9, Random(3-9)

Figure 3 Probabilistic Flooding: Parameter K vs Saved Rebroadcast.

A good reachability gives a worse SRB as with the case of AODV. But it can be noticed from the Figure 3 that a better SRB can be achieved with a random reachability parameter. If K is very less, many of the neighbors are chosen and only very few are blocked from rebroadcasting. Therefore when the value in K is 2, for the various network densities, the SRB is only 40 to 45%. If the value in K is very large, only very few of the neighbors are chosen for rebroadcasting and all other neighbors of the node are blocked. In that case, there is a chance that the destinations may be a neighbor to the blocked node. It may happen that, particular destination in this case may not be reached. Therefore the K parameter should neither be too small nor be too large. It is seen from the Figure 3 for K = 9, for the network densities of 50 and 75 the SRB are very less where as for the densities of 100 and 125 the SRB is good. Hence it is better to have a random value in K since there is no prior knowledge of the network size. The last set of bars shows the SRB for random value in K.

## 5.3 Path Optimality

In spite of reducing the number of control packets, the path established between the source and the destination is shortest in terms of number of hops as that of AODV. Each result in Figure 4 is the average of 50 source-destination pairs on top of 3 different network topologies for a given network density. For a smaller network size of 30, with a reachability parameter K = 2, average number of optimal paths achieved is very closer to AODV protocol which is arrived at by flooding through all nodes. It is clear from the graph that for larger networks the value of reachability parameter does not affect the optimal path much. Very large reachability parameter affects the path optimality. Larger the K values higher the number of neighbors blocked. Therefore even though there is a shortest route to the destination from the source, the packets take a longer (round about) path to the destination.

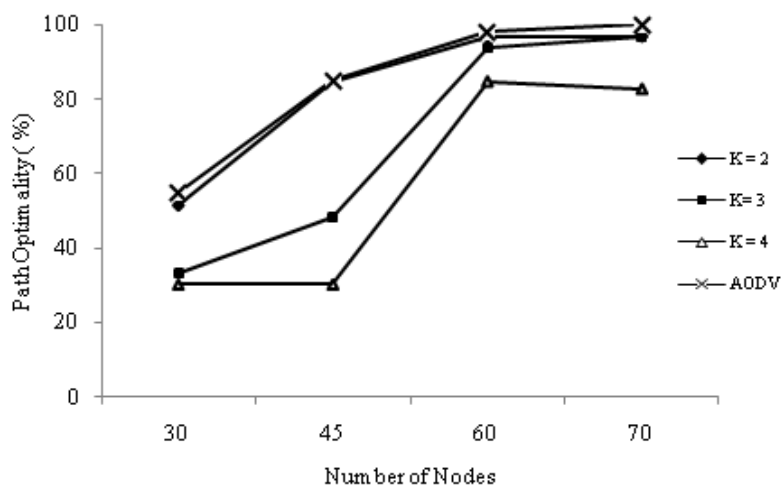

Figure 4. Reachability for various Node Densities vs Shortest Paths

In AODV, upon receiving the RREQ, each intermediate node checks whether it has an existing entry for the destination. If it has, a route reply (RREP) packet is generated and unicasted back to the source along the reverse path and thus the number of RREQ rebroadcasts are saved. But in the worst case, if none of the node have path to the destination, the RREQ packets are rebroadcasted till it reaches the immediate neighbor of the destination. Broadcast storm is caused when more number of nodes tries to establish communication path in dense MANETs. This leads to network congestion and hence there is consistent delay in finding route in AODV.

## 5.4 Impact of Node Density

Here we fix the traffic flow, the maximum node mobility to 20m/s and vary the number of nodes in the network to investigate its impact on AODV and PRP. The results are shown in Figure 5 in which each data point represents the average performance under 5 different topologies. It can be seen that route request success rate in both the protocols increases with the number of the nodes. With a small number of nodes, the network is not fully connected and so it is difficult to establish routes between a given source and destination. Due to lesser network connectivity and suppression on node from rebroadcasting, the route request success rate in PRP is very less compared to AODV. But, with the increased node density, it can be seen from the graph in Figure 5, the success rate is almost equal to the route request success rate of AODV.

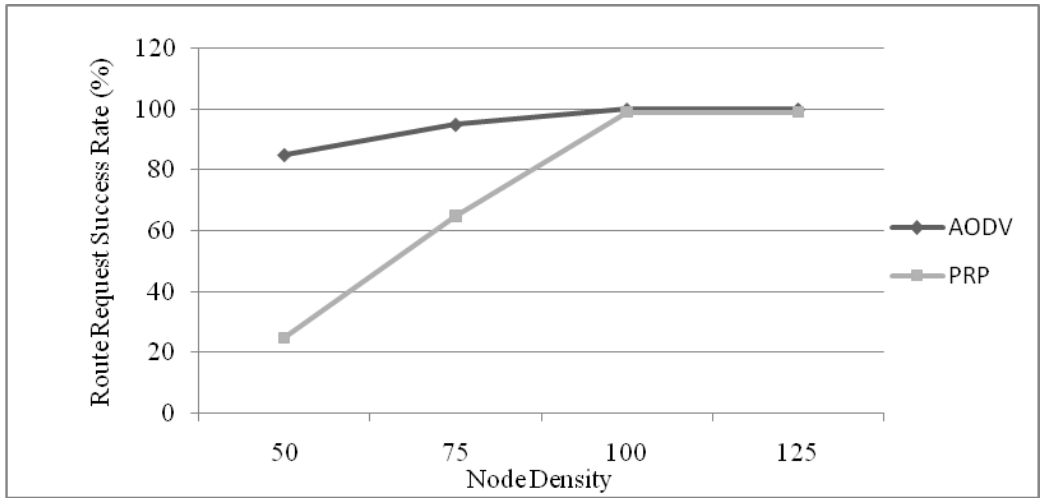

Figure 5. Relationship between the route success rate and node density

Figure 6 shows the impact of node density on SRB. From the graph in Figure 6 it is can be easily realized that our protocol drastically cut down a large amount of routing control overhead. Smaller control traffic translates to lower power consumption, less congestion, smaller delays, reduced memory and processing requirements and faster access to the communication channel.

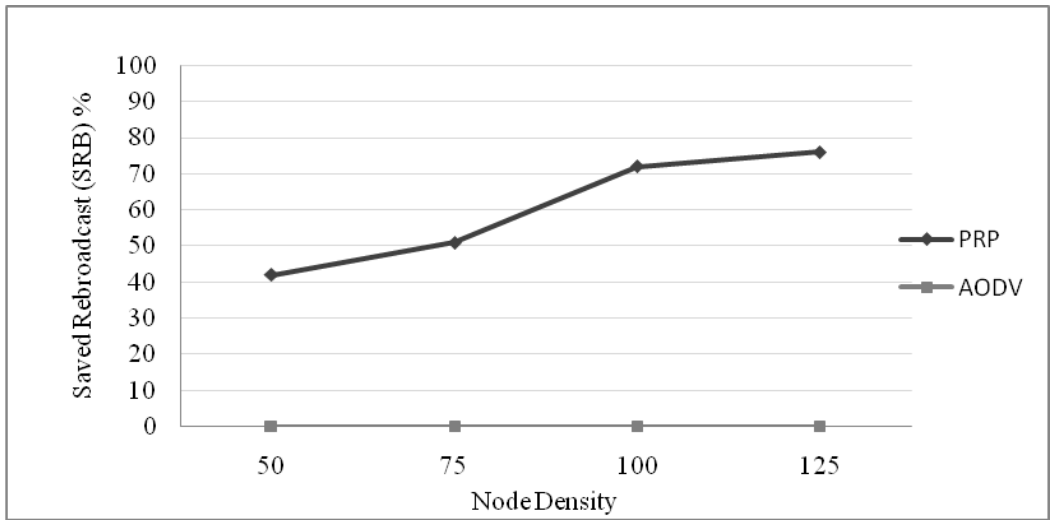

Figure 6. Relationship between the route success rate and node density

## 5.5 Impact of Traffic Density

Figure 7 shows the variation of SRB in AODV and PRP, where many different sources and destinations attempts to discover path simultaneously. The scenario is based on the density of the network of size 50, 75, 100 and 125 nodes. The number of active source and destinations are set as 2, 3 and 5. When the density increases the percentage of SRB also increases in PRP. On an average the 70% of rebroadcast is saved in PRP. This infers that the network is less congested even in case of many source-destination pairs trying to find routes.

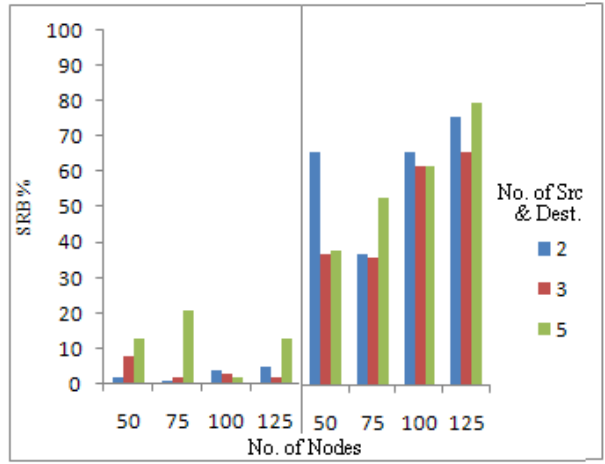

Figure7 Network Density vs Saved Rebroadcast for multiple Sources and Destinations

SRB depends on the following factors: the topology of the network, node density, and also the positions of the source and destination nodes in the network. (When there are just 50 nodes and only two pairs of source and destinations are trying to establish path between them the links are not more congested and the intermediate nodes in the paths are unique. So the number of RREQs generated are also very less. Nodes are not overloaded with more than two paths through it to avoid congestion during data transmission. The number of RREQs generated also depends on the topology of the network). It is evident from Figure 6 that PRP saves more number of rebroadcast as the density of the network increases and thus do not congest the network with RREQ packets even if multiple source and destination tries to discover path between them.

## 5.6 Path Acquisition Latency

Path acquisition latency is the delay involved in finding the path from the source to destination. The impact of node density and the traffic density on latency in AODV and PRP is shown in Figure 8. The latency involved in establishing 2, 3, 4 and 5 hops for the various node densities like 50, 75, 100 and 125 and various traffic flows is shown in the graph. It is clear that the latency involved in establishing paths of various lengths is same in PRP as compared to AODV.

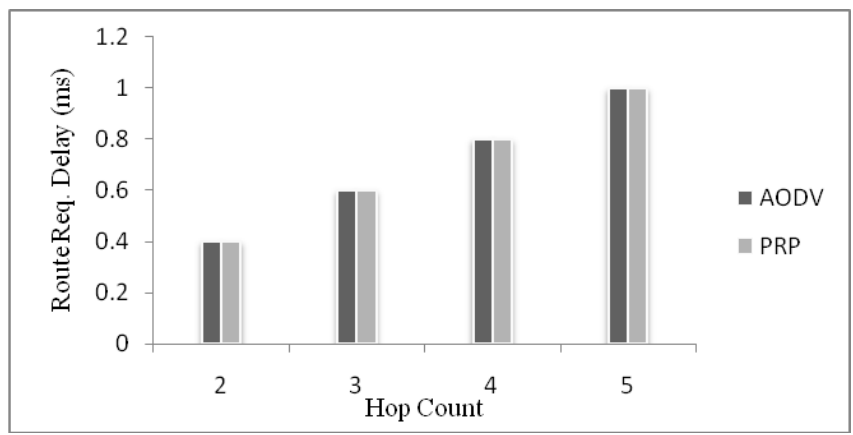

Figure 8. Route Request Latency

## 5. CONCLUSION

Larger density networks have large number of broadcast and congestion particularly when different sources initiates path discovery. This may choke the entire network and may cause packet drops due to which the path discovery may fail. In dense networks routing packets would consume the bandwidth usage rather than data transmission.

In this paper, we have proposed a new route discovery algorithm for mobile ad-hoc networks, with a reduced overhead in case of dense networks. It is particularly efficient in case of a high density of nodes. Our experiments have demonstrated, through analysis and simulations, a significant reduction in the number of rebroadcast messages. The number of rebroadcast saved almost reaches 95% when the node density is more. This reduction in the rebroadcast relieves the network from getting congested. Furthermore, because the nodes that rebroadcast the message are very close to the border of the radio area, the probability of getting an optimal distance is increased. But the latency and the success rate in finding the route are almost same as that of AODV (Figure 5 & Figure 8). Since our algorithm is based on probabilistic approach, there is small chance that the route request cannot reach the destination. Since in less dense networks, the Route Request Success rate is far from the optimal as compared with AODV, PRP do not adapt well to a networks with lesser node density. The proposed algorithm performs better than AODV in dense network.